# Interionic distance distributions between $Er^{3+}$ and $Yb^{3+}$ as dopants in some crystal matrix hosts used for upconversion luminescence

**Liviu Dudaș**

Faculty of Chemical Engineering and Biotechnologies, National University of Science and Technology Politehnica Bucharest, 1-7 Gheorghe Polizu Street, 011061 Bucharest, Romania; liviu_dudas@yahoo.com (L.D.);

**Abstract**: The upconversion process for the $Er^{3+}$ ion, when irradiated with IR photons at 980 nm, strongly depends upon the presence of the sensitizer $Yb^{3+}$ ions. There is a good correlation between the relative and absolute concentrations of the activator and sensitizer species and the intensities of the emission lines of the resulted visible spectrum. The ratios between emission intensities in the green band (510–580 nm) and red band (640–700 nm) (i.e., the spectral content) show similarities between different host crystals in which $Er^{3+}$ and $Yb^{3+}$ are embedded, which is an indication that, regardless of the crystalline medium of embedding, the dopant ions interact in some specific and similar ways. In order to reveal the mechanisms for the energy transfer between activator and sensitizer ions, one needs a model for the spatial distribution of these relative to one another, from which some statistical data, like crystal matrix average surrounding radii, could be inferred. Some models of cubic neighboring are presented for the relative concentration ratios of $Er^{3+}$:$Yb^{3+}$, respectively of 1:2, 1:4, and 1:8.

Keywords: interionic distance, upconversion, $Er^{3+}$, $Yb^{3+}$ …

## 1. Introduction

As seen in **Figure 1**, when illuminating with 980 nm some oxidic ceramic samples doped with $Er^{3+}$:$Yb^{3+}$, the more sensitizer $Yb^{3+}$ ions are present in the matrix, the more the red component (640 - 700 nm) of the upconversion emission of $Er^{3+}$ is enhanced while the green emission (510 - 580 nm) is diminished.

Since this behavior is constant across all the crystalline hosts studied by us [1, 2], regardless of their particular characteristics, like composition, density, crystal structure, symmetry of embedding site of $Er^{3+}$, dielectric constant, etc., it is most presumable that the parameter that governs this phenomenon is the medium distance between $Er^{3+}$ and $Yb^{3+}$ ions, whose value is, for most part at low concentrations, greater not only than the crystal's unit cell dimensions but even greater than the nanocrystal sizes that form the ceramic.

So, an evaluation of these average interionic distances, which depend on the concentrations of $Er^{3+}$ and $Yb^{3+}$, is mandatory for the understanding of the phenomenon of the aforementioned red shift. The effective measurement of these distances is impossible, at least for our possibilities, so the single way of assessing them, at least their order of magnitude, is the computer simulation.

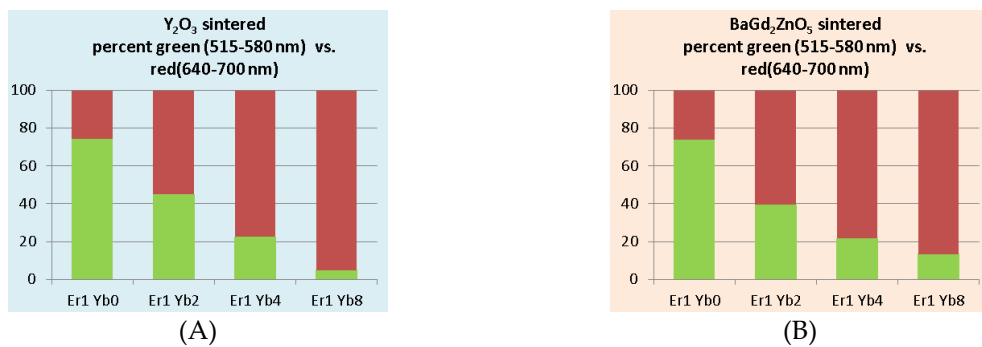

**Figure 1.** Variation of percentual green vs red emissions in the total visible emissions for the upconverted incident 980 nm radiation which illuminate and (A) $Y_2O_3$ and (B) $BaGd_2ZnO_5$ ceramic samples doped with $Er^{3+}$:$Yb^{3+}$ with the respective (horizontal axis) absolute ratios relative to the substituted ion ($Gd^{3+}$, $Y^{3+}$).

## 2. Simulation setup

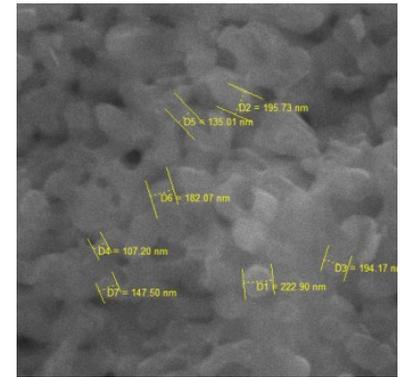

Since the mixing of the dopants into the crystalline matrix is assumed to be totally random, i.e., no segregation, no gradients, no periodicity, no correlation, one must start with a statistically total homogeneous situation.

As a starting point, it was considered the average dimensions of a nanocrystallite, which are revealed by SEM investigations. For example, in the image at right, the crystallite dimensions are in the range of 100-200 nm.

This is why, for the abstract sample used in the simulation, a cube with 400 nm edges was considered appropriate.

For each case of the crystalline matrix considered ($BaGd_2ZnO_5$, $Y_2O_3$, etc.), its crystallographic data (CIF files) from [4] was taken into account, and the 400 nm cube was filled with unit cells, packed by translation on each axis with the appropriate unit cell dimensions ($a, b, c$) constant. Since the compounds used had, all of them, rectangular unit cells (cubic, orthorhombic), this procedure was simple.

The positions, in each unit cell, of the substitued ions ($Y^{3+}$, $Gd^{3+}$, etc.) were recorded in a list, and, by using a random number generator that generates constant (flat) distributions, the appropriate number of dopant ions for each concentration case (%$Er^{3+}$=a, %$Yb^{3+}$=b) were distributed.

The second step was to calculate, for each $Er^{3+}$ random position, the distances to all the other dopants, either the same kind or the sensitizer one ($Er^{3+} \leftrightarrow Er^{3+}$ or $Er^{3+} \leftrightarrow Yb^{3+}$). The results were put in a list, and this list was ordered from the smallest to the greatest.

The next step was to take, from each list (there are so many lists as kinds of dopants are), the first n = 1, 2, 4, 6, 8 values, make their arithmetic average, and build another list (this time is only one list, and its length is equal to the number of dopants) with these averages.

This averages list was also ordered increasingly, and the number of values that were situated in fixed intervals were counted (e.g., count all distances with values in the range 4.0 Å ≤ $v$ < 5.0 Å).

This final list was the distribution that was looked for and which assigns, to each interval of distance (usually 1 Å in length), the number of dopant ions that have the closest *n* neighbors to the average distance $R_n$ in that interval.

A suggestive 2D drawing that helps to understand the computation is presented in **Figure 2**.

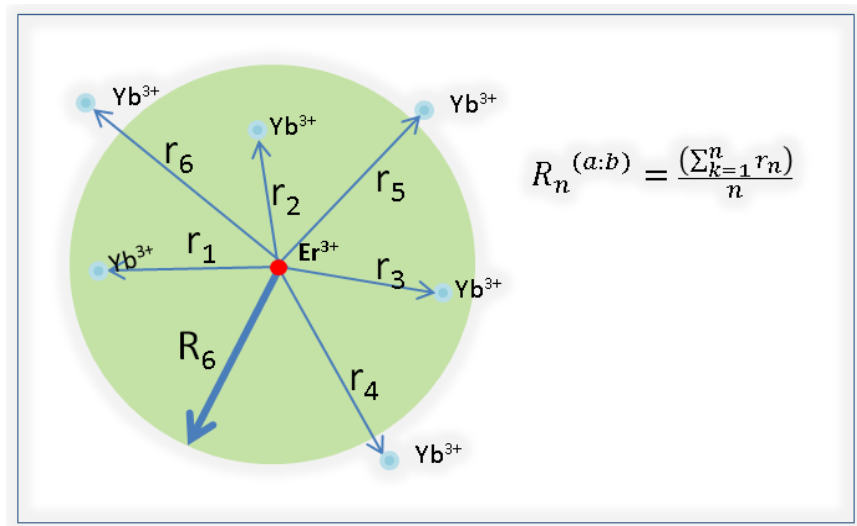

**Figure 2**. 2D drawing suggesting how the average radius $R_n^{(a:b)}$, of the cloud of $n$ Yb³⁺ ions surrounding an Er³⁺ ion for %Er³⁺:%Yb³⁺=(a:b), was considered.

## 3. Simulation results

**Figures 3** and **4** show the calculated distributions for two compounds, $Y_2O_3$ and $BaGd_2ZnO_5$. The curves are for each specified case of *n* $Yb^{3+}$ neighbors for 1:2, 1:4, and 1:8 relative $Er^{3+}$:$Yb^{3+}$ percentual molar concentrations.

Observe how increasing the concentration of $Yb^{3+}$ results in shifting the distributions toward lower values, i.e., each $Er^{3+}$ is more surrounded by $Yb^{3+}$, which, of course, was expected. The vertical axis in each graph represents the percentage of $Yb^{3+}$ ions that are at the distance specified by the horizontal axis.

Graphs show the distribution of distances between $Er^{3+}$ and $Yb^{3+}$ ions in the following cases:
- percentage of distance to the nearest neighbor (zigzagged dotted lines). One sees that the distribution for the closest neighbor is not linear because of the special positions that can be occupied by the dopants in the crystal matrix formed by specific unit cells.
- percentage of the average distances of the nearest 4, 6,…, 32 $Yb^{3+}$ neighbors. As expected, the distributions are Poisson-like [5], but each with different parameters, which come from the 3D spatial distributions and the specificity of the starting CIF data.

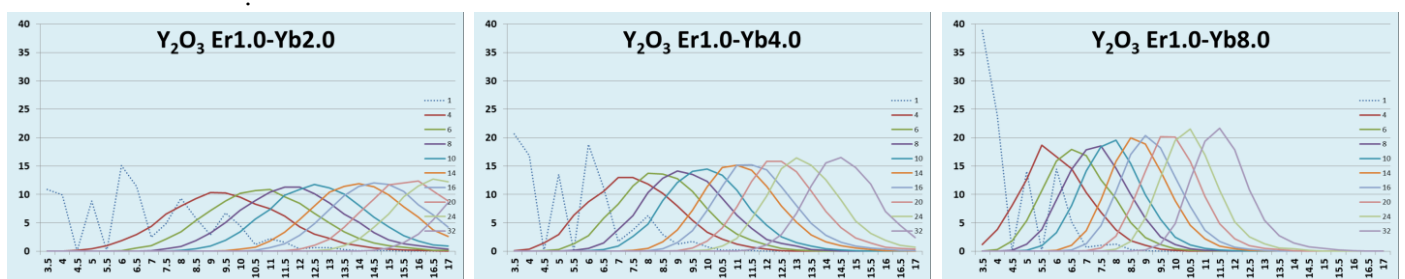

**Figure 3**. The distance distributions for $Er^{3+} \leftrightarrow Yb^{3+}$ when embedded in $Y_2O_3$. Observe, in the case of Er:Yb=1:8, almost 40% of $Er^{3+}$ ions have the closest $Yb^{3+}$ sensitiser at 3.5 Å from themselves. (y-axis represents the percent of ions having *n* neighbours at $R_n$=x Å)

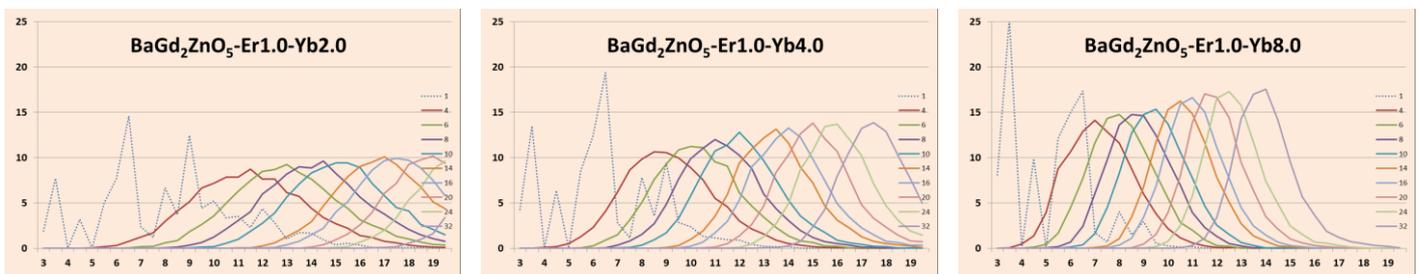

**Figure 4**. The distance distributions for $Er^{3+} \leftrightarrow Yb^{3+}$ when embedded in $BaGd_2ZnO_5$. Observe, in the case of Er:Yb=1:8, almost 25% of $Er^{3+}$ ions have the closest $Yb^{3+}$ sensitiser at 3.5 Å but around 8% of $Er^{3+}$ have the $Yb^{3+}$ sensitisers even closer, at 3 Å. (y-axis represents the percent of ions having *n* neighbours at $R_n$=x Å)

## 4. Discussion

If one takes the abscissa value of a distribution peak as the value for the medium radius of encompassing for the respective number of neighbors, we see that, for the relative $Er^{3+}$-$Yb^{3+}$ concentration ratios of 1:2, 1:4, 1:8, certain average radii are equal, i.e., $R_4^{(1:2)} = R_8^{(1:4)} = R_{16}^{(1:8)}$, and $R_6^{(1:2)} = R_{12}^{(1:4)} = R_{24}^{(1:8)}$.

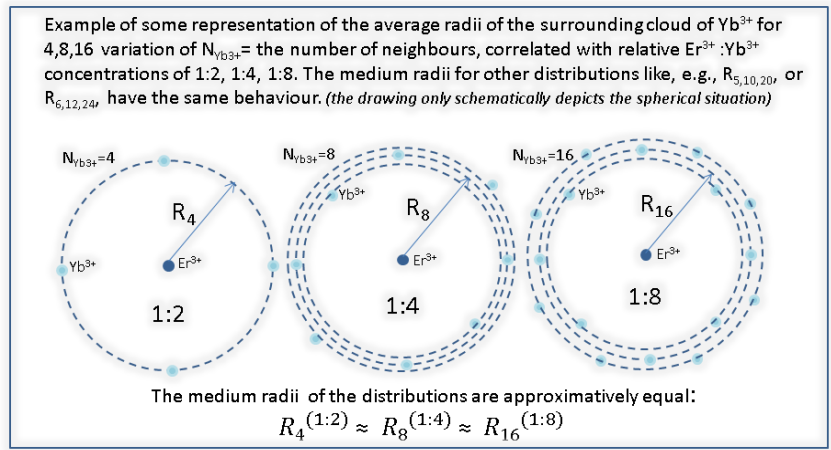

**Figure 5**. Simulation results indicating what concentrations have, for their peaks of the distributions, the same average radius $R_n^{(a:b)}$.

Observing where the peaks of the distributions of distances are situated for the different concentrations, for the two cases of crystal hosts discussed above, one can observe that the average values for $R_4$, $R_8$, and $R_{16}$ are those in **Table 1**.

**Table 1**. By observing the distributions in **Figures 3** and **4**, the peak values $R_n^{(Er3+:Yb3+)}$ for $R_4^{(1:2)}, R_8^{(1:4)}, R_{16}^{(1:8)}$, can be extracted.

| Host crystal → <br> Radius ↓ | $Y_2O_3$ | $BaGd_2ZnO_5$ |
|---|---|---|
| $\langle R_{4,8,16} \rangle$ | 9.25 Å | 11 Å |
| $\langle R_{6,12,24} \rangle$ | 10.25 Å | 12.5 Å |

Starting from the distances estimated from the above graphs, some simple surrounding models can be built in order to further estimate the mechanisms for the energy transfer between activator and sensitizer ions.

Some models of neighboring are considered for the relative number ratios of $Er^{3+}:Yb^{3+}$, respectively of 1:2, 1:4, and 1:8, and are shown in **Figure 4**.

The principle of choosing the neighboring unit cell is the maximal distribution of distances between $Er^{3+}$ in the center and the surrounding $Yb^{3+}$, so, in the case of 1:4 where two cases of unit cells exist, the preferred one is that with $Yb^{3+}$ ions on faces and not on edges, since the $Yb^{3+}$ is more uniformly distributed.

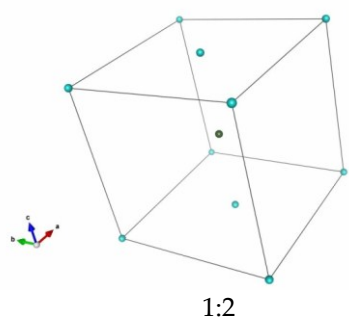

1:2

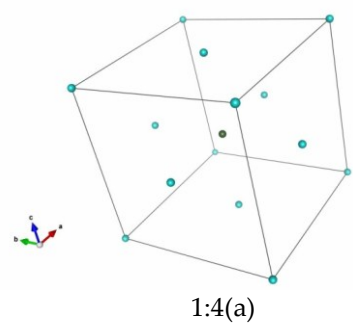

1:4(a)

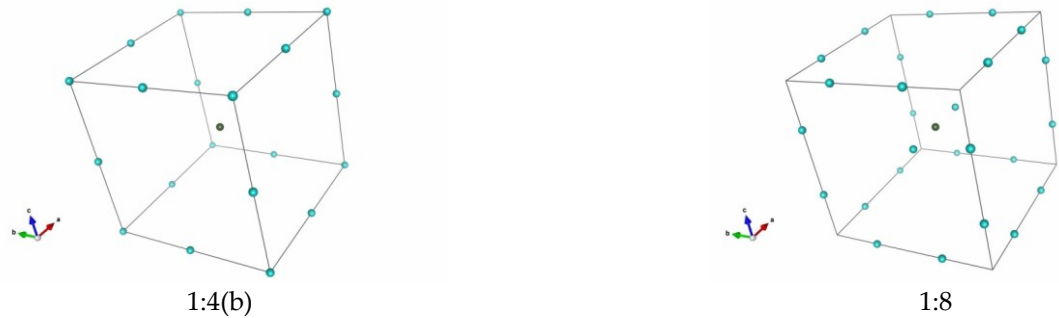

**Figure 4**. Some simplified cubic models of surrounding a central Er$^{3+}$ ion by 2,4,8 Yb$^{3+}$ sensitisers. The placement schemas can be used to further build some models of Yb$^{3+}$ cavities and energy interactions between activators and sensitisers. Central Er$^{3+}$ ion with dark grey, Yb$^{3+}$ ions with cyan. The cube edge in the images is 18.6 Å, in order to better see the relative Er-Yb positioning, otherwise the edge lengths should have had the values of the peaks of the distributions.

## 5. Conclusion

The above simulations give important indications why the red (640–700 nm)/green (510–580 nm) ratio intensities are in direct link with the number of Yb$^{3+}$ sensitizers on a fixed-radius sphere, which shows that, at least at small concentrations, Er$^{3+}$ ions interact with the neighboring Yb$^{3+}$ ions as these form a kind of mirror with certain degrees of reflectivity.

Regarding the phononic interactions, these should be the same, regardless of the concentration. This is because the phononic band structure of the host crystal isn't very strongly perturbed by the dopant ions since they are well replacing the base ions (Gd$^{3+}$ or Y$^{3+}$) in the lattice and have comparable masses with these.

Another reason is that the concentrations of dopants are sufficiently small and do not induce lattice deformations or structural changes.

Comparing the values in **Table 1**, one sees that in the case of BaGd$_2$ZnO$_5$, the considered $R_4^{(1:2)}, R_8^{(1:4)}, R_{16}^{(1:8)}$, radii are 20% larger than in the case of Y$_2$O$_3$, yet the efficiency of the former is higher. This shows that, beside the average distances, for assessing the absolute value of the upconversion efficiency, the particular parameters of each case of the crystalline host should be taken into account (like phonon frequencies or dielectric constant).